# Emerging Networks and Services in Developing Nations- Barbados Use Case

Warren Scantlebury
psxws4@exmail.nottingham.ac.uk
School of Computer Science
The University of Nottingham

Milena Radenkovic
milena.radenkovic@nottingham.ac.uk
School of Computer Science
The University of Nottingham

## 1. Introduction

There is very little research done in the context of developing countries and the performance of various DTN routing protocols. Compounding this problem is that the DTN research is still a relatively new field and is therefore more likely to be tested in more developed countries. One reason is that many of the infrastructural systems that are often desired to complement DTN research use cases such as 5G networks or GTFS feeds for public transport schedules are not available in developing countries making them less appealing to researchers. In addition larger universities and institutions in developed countries will be more likely to have the resources and funding available to facilitate pioneering research. Even though Barbados is a relatively small island the public transport system has been known to be very inefficient with many bus routes being under-served due to financial and reliability issues and has in the past resulted in the Transport Board explicitly warning users of long wait times [23]. These problems can subsequently lead to crowded buses which bring serious discomfort and pose many safety risks to passengers, as accidents involving public transport vehicles are a common occurrence[10][11][22]. Additionally, the Transport Board has also been operating at a significant loss, being reported to be the primary source of losses among state owned enterprises by the ministry of Finance [15], with an ex-chairman even testifying to have used his personal credit card to purchase equipment while overseas, in order to save money for the organisation [16]. We are therefore seeking to not only investigate DTN performance on a map of Barbados, but to additionally explore whether DTN communication can provide a viable means by which the efficiency of the public transport system can be improved.

This report aims to conduct an in-depth comparison of DTN (Delay/Disconnection Tolerant Network) performance and characteristics in the developing country of Barbados versus two major UK cities Nottingham and London. We aim to detect any common patterns or deviations between the two region areas and use the results of our network simulations to draw well-founded conclusions on the reasons for these similarities and differences. In the end we hope to be able to assimilate specific portions of the island to these major cities in regard to DTN characteristics. We also want to investigate the viability of DTN use in the transport sector which has struggled from a range of issues related to efficiency and finance, by recording and analysing the same metrics for a DTN that consists of only buses.

This work is intended to serve as a bridge for expanding the breadth of research done on developed countries allowing other researchers to be able to make well informed assumptions about how that research may apply to developing nations. It will consist of results that show graphical trends and analysis of why these trends might exist and how they apply to real world scenarios.

## 2. Barbados – Background, Motivation Related Work

### 2.1 Barbados overview

Barbados is a small developing island nation in the eastern Caribbean. It is a popular tourist destination, and unlike many of its close neighbours wasn't formed by volcanic activity. The island has a tiny footprint of about 430km2 and is home to roughly 290,000 people [4]. One of the major campuses of the region's University of the West Indies is situated on the island attracting thousands of students from across the region every year. Barbados is generally described as relatively flat when compared to many of its Caribbean neighbours, with its highest point being at the peak of Mount Hillaby at 340m. However its landscape is far from featureless as Mount Hillaby is accompanied by the well-known Scotland District [19], the only part of an underwater mountain range also known as an accretionary prism [6, 24] that reaches above sea level and a portion of the island that among many other things is prone to erosion, due to its hilly nature and the composition of its soil.

#### 2.1.1. Barbados transport board

The Barbados Transport Board is a state owned agency that provides public transport for residents of the island and covers most of the island with its roughly 70+ routes, managing the scheduling and maintenance of a fleet of around 170 buses. The agency operates out of three main terminals, two located in the capital of Bridgetown, (Fairchild Street and Princess Alice Highway) and one located in the northern town of Speightstown. These terminals roughly split the island into 3 sections with the Fairchild Street terminal serving primarily the central and south-eastern portions of the country, the Princess Alice terminal serving the western and some central parts of the island and the Speightstown terminal serving the northern parts of the island. There is some overlap between terminals in order to make sure the island is well connected, for instance there are several routes that leave the Princess Alice terminal destined for the northern regions of the island, often making a stop at the Speightstown terminal, similarly there are routes that connect the southern parts of the island with the terminal in Speightstown.

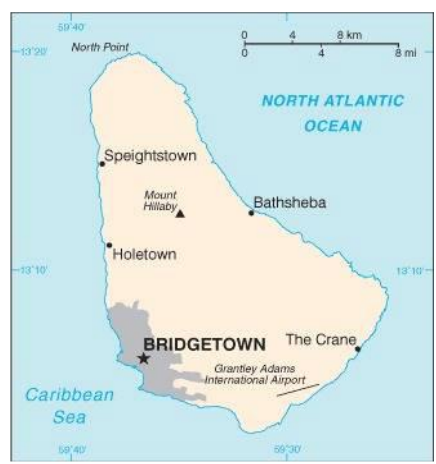

*Figure 2.1 Map of Barbados*

#### 2.1.2 Barbados Public service vehicles (PSV)

The Barbados Transport Board is by and large mirrored in its operations by Privately owned public service vehicles (PSVs), these PSVs are given special licences to operate on certain routes across the island, the majority of which are identical to those in operation by the Transport Board, both owned and operated by private citizens and businesses these PSVs fill a crucial gap and account for over 70% of all public transport vehicles in operation. Because there is no one organisation overseeing their daily operations, each PSV tends to operate in a manner that will benefit it individually. The result is that there is no set schedule and no predefined start and end time for a PSV, on routes with a high number of PSVs the effects of this unpredictability on waiting commuters will be largely mitigated due the fact that there is always PSV within a certain distance of each bus stop.

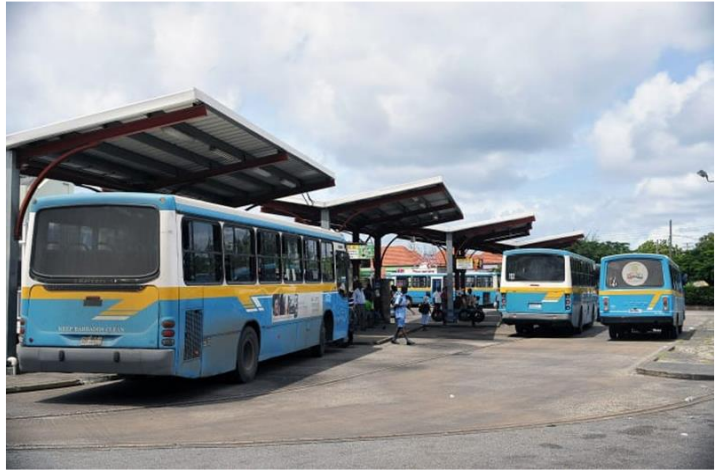

*Figure 2.2 Transport Board buses being boarded in the Transport Board section of the Princess Alice terminal*

### 2.1.3 Barbados transport authority

Overseeing the operations of all public transport in Barbados is the government run Barbados Transport Authority (TA). While the transport board and each PSV owner are in charge of the finer details of their own operations, the TA sets the guidelines within which each entity must operate. Some of the more significant responsibilities of the TA include the designation of specific routes for buses and PSVs to follow, and specifically for privately owned PSVs the TA issues licences to not only sanction the operation of each vehicle as a PSV but to also determine which route it is allowed to operate, compared to the Transport Board whose fleet of buses can operate on any route they service. Figure 2.4 shows the PSV permit activity in an annual report covering April 2019 to March 2020, The B and ZR columns represent privately owned PSVs while the BT column represents buses owned by the Transport Board, other columns represent other classification of PSVs such as Z representing taxis for hire and H for hired cars. The active permits on this table show how privately owned PSV's outnumbered the Transport Board buses by a ratio of 5:1, and approved applications were more than 6 times higher for privately owned PSVs.

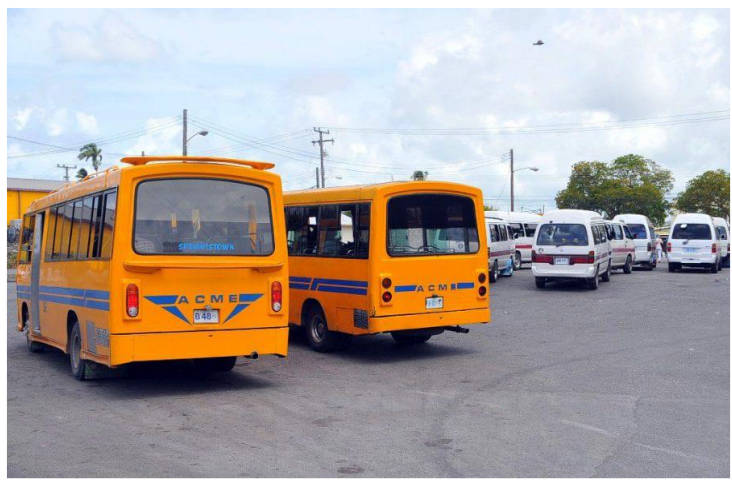

*Figure 2.3 PSV vehicles awaiting passengers in the PSV section of the Princess Alice terminal*

| CATEGORY | Z | ZM | ZR | H | BT | B | BC | HL | HC | TOTAL |
|---|---|---|---|---|---|---|---|---|---|---|
| Applications Received | 334 | 318 | 378 | 1044 | 87 | 210 | 2 | 6 | 18 | 2397 |
| Applications Approved | 67 | 60 | 84 | 596 | 22 | 65 | 3 | 0 | 19 | 916 |
| Permit Transfer Request | 17 | 4 | 10 | 33 | 1 | 7 | 0 | 0 | 0 | 72 |
| Permit Transfers Approved | 9 | 1 | 6 | 24 | 1 | 1 | 0 | 0 | 0 | 42 |
| Imposts from Transfers | $22,500 | $2,500 | $6,500 | $84,000 | $16,000 | $12,000 | 0 | 0 | 0 | $176,000 |
| Active Permits | 2229 | 834 | 502 | 3860 | 148 | 241 | 77 | 43 | 21 | 8092 |
| New Activations | 37 | 38 | 23 | 199 | 13 | 15 | 0 | 1 | 0 | 404 |
| Imposts from Activations | $92,500 | $133,000 | $149,500 | $696,500 | $208,000 | $12,000 | $0 | $4,000 | $0 | $1,541,500 |

*Figure 2.4: PSV applications received during a one year period from 2019-2020*

### 2.1.4 Barbados Telecommunication services

On the island there are two major network providers, Flow and Digicel who provide both mobile and fixed line internet service to the island. As it relates to both carriers, the island typically has good cell coverage in more urban and populated areas, but rural areas often have poor reception. All 11 parishes have access to their landline and home internet services, and the two network providers also have their own cable TV offerings, also available island wide. The cost of a mobile data plan is relatively expensive per GB, and plans often have very short time limits such as 7 or 14 days, forcing users to renew or lose their data. From Table 2.1 we can see a comparison of Flow and Digicel's prepaid plans for 7, 14 and 30 days and how the price per GB varies for different plan lengths. At the time of this research the Digicel prepaid plans were much cheaper and significantly more stable than flow, varying by less than 15% while Flow saw an increase of 80% from the 7 to 30 day plans. Compared to the UK offerings in Table 2.2 these plans are both expensive per GB and also structured in such a way to prevent low data users from purchasing small amounts of data for longer periods such as a month as both UK offerings have 30 day plans for all data amounts.

*Table 2.1: The price per GB for different Flow and Digicel data limited plans*

|  | 7 days | 14 days | 30 days |
|---|---|---|---|
| Digicel | 6 | 6.85 | 6.67 |
| Flow | 12.50 | 15 | 22.50 |

*Table 2.2: The price per GB converted to Barbados Dollars for different 30 day data limited plans*

|  | $23 | $35 | $47 |
|---|---|---|---|
| Smarty | $0.77 | $0.70 | Unlimited |
| Vodafone | $1.10 | $0.86 | $0.25 |

## 2.2 Delay/Disconnection Tolerant Networks (DTN)

DTNs first described in [8] refer to a type of network architecture that is designed to operate in "challenged" environments such as those prone to disconnections, or high delays. These challenges faced can be as a result of the network infrastructure or other characteristics of the network such as node mobility or performance. The range of scenarios that can create challenged networks are diverse and unique and can range from military combat, to satellite/interplanetary networks [7, 17], to a post disaster scenario [2, 3]. While the aforementioned scenarios can all be intuitively linked to a poor performing network, there are other scenarios that may be less obvious, for example a crowded event where everyone is using their phone to communicate, this can quickly become a situation where the network is quickly overcrowded and becomes slow and even unresponsive. In DTNs, nodes within the network have the ability to self organise and facilitate communication without the need for any central infrastructure or authority and integral to this self organisation are the routing protocols. Routing protocols are the set of rules that determine the procedure to be followed when sending, storing and forwarding messages, and determine what steps a node needs to take in order to get the messages in its care to their recipients. In addition to enabling autonomy, routing protocols play a big part in determining how a DTN will perform [1, 13, 20] as these networks have highly variable topologies. From a very high level there two types of routing protocols, flooding protocols which relies on spreading multiple copies of messages into the network to increase its chances of delivery and forwarding protocols which will focus more on getting the original copy of the message to its destination. Many recent protocols will typically have elements of both flooding and forwarding.

Probabilistic Routing Protocol using History of Encounters and Transitivity (PROPHET) [14] is a routing protocol which utilises flooding in a more controlled way compared to other flooding protocols. In PROPHET the notion of delivery predictability is introduced indicating the likelihood of a message being delivered to its destination by its current host. This delivery predictability is calculated for each node and recalculated on subsequent contacts. PROPHET takes into account how recently two nodes have seen each other in order to ensure that nodes who have not come into contact for a long period of time will have a lower probability of delivering messages to one another. This "aging" of the delivery probability can be configured via a time constant k as seen in the following equation $P(a,b) = P(a,b)_{old} y^k$ making it possible to age messages by only a fraction of the time that has passed. A transitive property is also included in the delivery predictability with the reasoning that "if node A frequently encounters node B, and node B frequently encounters node C, then node C probably is a good node to forward messages destined for node A". When forwarding messages only nodes with a higher delivery probability for the destination will receive those messages and the current node will retain that message until it needs to free buffer space for newer messages, this is to ensure that if a node was to later come across another contact with an even better delivery probability it can forward to that new contact.

The Spray and wait protocol described in [25] is a routing protocol that attempts to increase the efficiency of typical flooding protocols by "spraying" a number of copies of a message before stopping. SaW does not require any knowledge of the network to operate. At a high level there are two phases to this protocol the "spray" phase where a generated message will have L copies eventually forwarded to L different nodes, and the "wait" phase where after all L copies have been fully distributed the nodes in possession carry out direct transmission and forward the message directly to the recipient. These two phases turn SaW into what is essentially a hybrid between flooding and forwarding protocols. The distribution of copies in the spray phase can either be done in "source" mode where the source of the message forwards a single copy to the first L nodes it contacts, or "binary" mode where the source and any other node that has more than one copy (n) available will forward n/2 copies to every node it meets until it has only one copy left.

## 3. Barbados Use Case Experiments Design and Setup - Maps

For the assessment of our scenario we decided to choose the ONE simulator [12] and integrate Barbados map from MyGeodata Cloud [18] an online geographical data processing application that allows for the extraction of map data from OpenStreetMap [5] and conversion into a format usable by the simulator. For each of the map locations used, an area of about 80km2 was extracted in order to ensure that differences in network performance and characteristics could be attributed directly to variations in landscapes and road infrastructure. Maps of different sizes are likely to have a significant impact on network characteristics as node densities and contact times will likely be higher since the same number of nodes will be contained in a smaller space. The coordinate system used was in the British National Grid format (EPSG 27700), which defines how the coordinate values are interpreted and all map files were in the well know text (WKT) format. When extracting road data from OpenStreetMap many of the paths specified were classified as "tracks" which upon further inspection were shown to be comprised of walkways carved out by repeated human activity, or gravel roads that gave access to large areas of private property such as farm land. Because these paths would have been used very rarely by the general public a conversion script was created to separate this data from the established roadways that would be used during the simulations. The conversion script also converted the file from the CSV format obtained from online to WKT.

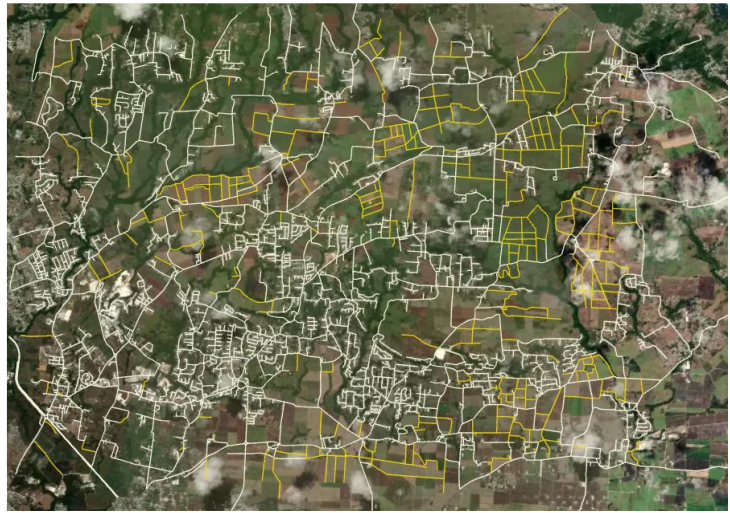

*Figure 3.1: Showing the mapping of the center of Barbados with extracted tracks in yellow and normal roads in white*

The areas of Barbados selected for experimentation were chosen in order to give as much insight as possible into the type of network topology created across the island and to analyse what similarities could be drawn from bigger, more developed cities. Identical experiments were run on the northern, south-eastern, central, and south-western parts of the island, the latter of which contains the capital of Bridgetown, and is the most densely populated part of the island. In order to have a good base for comparison, sections of two British cities Nottingham and London were extracted giving the ability to compare performance and characteristics with cities in developed versus developing countries. In the case of Barbados we can see how the road density varies greatly from the map of the capital shown in Figure 3.2 compared to that of the map of the north in Figure 5.11 and the center of the island in Figure 3.1, highlighting the variation in road density seen in a country merely 430km2 in size. In addition to the maps extracted from the 4 areas of the island there was a fifth map containing the paths of 8 of the bus routes that departed from the Princess Alice terminal and overlapped at various different points.

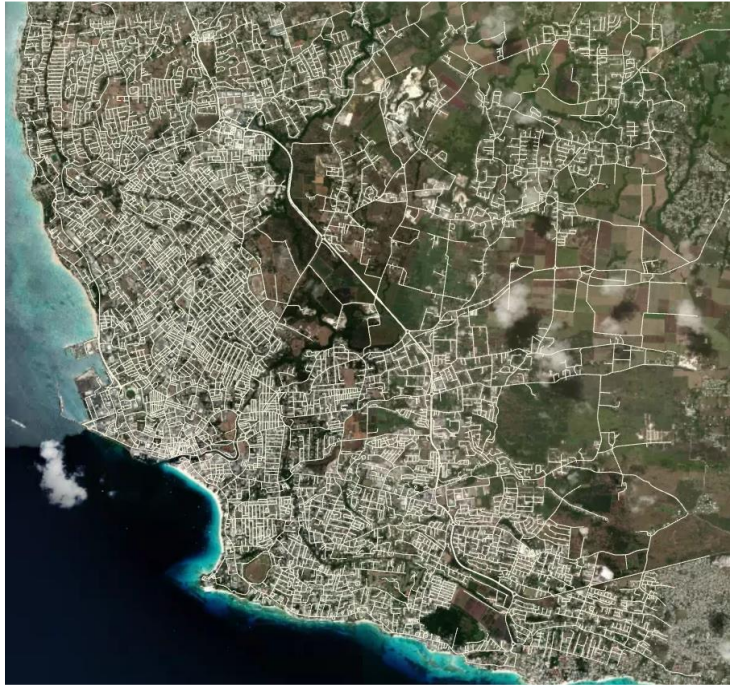

*Figure 3.2: Showing the roads extracted from the capital of Barbados and its surrounding areas*

The experiment was run with two different types of nodes, vehicular and pedestrian with different speed of travel. All simulated communication was done via Bluetooth radio and all nodes used the same routing protocol per simulation, meaning that any two nodes in contact could exchange information. In this report we investigated a number of performance factors, namely Bluetooth range, Bluetooth speed and buffer size, frequency of messages and number of nodes. The same combinations of variables were tested for two different routing protocols PROPHET and SaW and for all map configurations, while being repeated with 3 different seeds for the random number generator ensuring different node locations and travel patterns. Each simulation run generated a number of out of the box reports from the ONE simulator, which were then averaged across all random number generators and used to compare and contrast the key performance metrics of the networks. For the type of movement model used where random travel destinations are set for nodes to travel to along the path of the map, we expect poor performance from a prediction based protocol such as PROPHET as it is design for scenarios where node movement is more life-like and habitual such as travelling to and from a workplace during the day and returning home at night time, we also expect that SaW should fare much better. However when we later introduce the scenarios containing bus routes the movement of each individual bus along one select path per bus should create an opportunity for the delivery predictability of PROPHET to be more useful.

For the routing protocols used there were some configuration options unrelated to the topology of the network that could have significant impact on performance characteristics. In order to determine the values that would be used throughout the remaining simulations for these configuration options we investigated the performance of a range of values for these parameters. For the SaW router this involved running the same simulation for all combinations of binary and source SaW with message copies ranging from 2 to 32 messages and choosing which combination gave the best performance. For the PROPHET router the only configuration parameter was the amount of seconds that made up a time unit when ageing the delivery predictions and the simulations were run for different values between 1 and 70.

# 4. Barbados Use Case Experiment Design and Setup - Routes and Movement

In order to obtain the paths specifying bus routes in the simulation, the route information posted on the website of the Barbados Transport Authority was used to trace the paths of 8 different bus routes using open jump, and each route was then extracted into its own WKT file to be used in the simulator. Some aspects of the ONE Simulator had to be modified in order to ensure the node placement and movement for bus routes was more practical, this involved making modifications to how nodes were placed and their initial direction along the route. Ideally bus nodes should be evenly spread along the route in order to minimise wait times for passengers. The code snippet in Listing 4.1 below shows how we modified the constructor of the movement model to calculate the number of stops that should be between each pair of nodes and also determine the direction of travel.

```
int nfOfHosts = settings.getInt( NR_OF_HOSTS );
// if we only have 1 stop then the equation results in divide by 0
if( nfOfHosts == 1) {
      step = route.getNrofStops() - 1;
} else if( nfOfHosts > 0){
      step = route.getNrofStops()/nfOfHosts;
}
if  (  settings.contains  (  ROUTE_DIRECTION_S  )  &&  settings.getSetting
(ROUTE_DIRECTION_S).equals("reverse")) {
      this.reverse = true ;
}
```

*Listing 4.1: A code snippet showing the number of steps being calculated and checking for the reverse direction setting*

We check to see if we have multiple hosts and then subsequently calculate a step value that would be used to increment the first stop index for each node, ensuring all nodes had even space between them. If the index for the first stop was calculated as being larger than the total number of stops then we simply reset the first stop index to 0. In order to extract the bus routes used in this experiment there was a fifth portion of Barbados extracted from OpenStreetMap that covered the west coast pictured in Figure 4.1. Each route was then extracted line by line and placed into a separate file before being merged in order to create one continuous route for the simulator, Figure 4.2 shows the paths of the different routes used and how they overlap with each other as they travel along the west coast of the island.

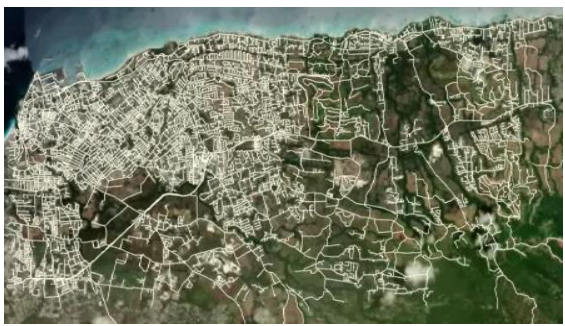

*Figure 4.1: Map of the west coast Barbados containing the roads from which routes were extracted (rotated 90◦ clockwise)*

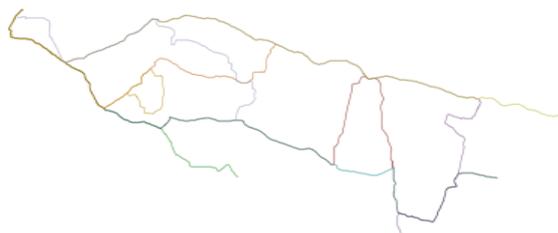

*Figure 4.2: image showing the paths of all 8 bus routes and how they overlap (rotated 90◦ clockwise)*

In order to achieve the separation between tracks and roads mentioned in Section 3 a Node.js script was written that read the column specifying path type and only retain those not listed as tracks.

# 5. Evaluation

The first stage of evaluation involved selecting the optimum routing protocol parameters for the given scenarios, this was done by analysing the results of the parameter optimisation simulations and selecting the configuration that gave the best trade off between resource usage and performance. After the parameters for each routing protocol were selected then the results from across the runs with different random number generator seeds were combined using the arithmetic mean, with one exception being the distance delay report where, across all the reports all values were combined into one file.

In order to carry out the performance evaluation of our various simulation scenarios, we have defined the following metrics which will be calculated and compared for all scenarios. Average Latency: the average time elapsed from when a message is created, to when it is delivered to its destination; Created: the total number of unique messages created during a simulation run; Started: the number of times two nodes have started the transfer of a message from one to the other; Relayed: the number of message transfers successfully completed, not to be confused with delivered messages that refers to those that have been delivered to the destination; Aborted: the number of message transfers that have been aborted; Dropped: the number of messages dropped from the buffer; Delivery probability: the probability of a message being delivered for a scenario (created/delivered); Overhead ratio: the ratio of how many additionally transfers were made when compared to the total number of messages delivered; Average hop count: the average number of hops required to deliver a message; Average buffer time: the average amount of time a message is spending in the buffer of an individual node. In addition to these metrics we will also analyse buffer occupancy reports which give a timeline of the average buffer occupancy for all nodes throughout the simulation, contact time reports which aggregates contacts by how long they last to the nearest second, and distance delay reports which reported the distance of every new message from its destination, how long it took to reach the destination and how many hops were required.

## 5.1. Parameters Selection

In our simulations that investigated the performance of different time units we saw that smaller increments of time unit were usually performing as good as, or better than larger increments, except in a few instances where it is common to see more reliable networks produce metrics that appear to be worse such as with increased hop counts and buffer times as a result of more messages with less straightforward paths to their destination being delivered. As shown in Figure 5.1 in those cases it tended to be the smallest value of 1 second consistently produced the best results, after which the performance saw a gradual decline before remaining constant, this behaviour was consistent across all maps. Our analysis of the results show that a 60% increase in message transfers started when moving from source to vanilla SaW resulted in a cascading effect that meant source SaW had a significantly lower resource usage, resulting in less than 50% of the amount aborted message transfers. This significant reduction in resource usage had no noticeable negative effect on performance as the delivery probabilities and average latencies for both versions were both very close. The data also reveal that the source SaW has a tendency to cause map performance to diverge into two distinct groups, giving the first insight that there are potential similarities between the Barbados north and Nottingham maps. When increasing the number of copies the delivery probability saw an initial rise before tending to gradually decrease as copies continued to rise, the London map had the slowest rate of decrease after reaching its initial peak, while the map covering the north of Barbados had the highest rate of decrease. The average latency declined sharply as copies were increased up to 8 copies before the rate of decline slowly tapered off. Unsurprisingly, metrics that dealt more closely with resource usage all got proportionally worse as the number of copies increased, this included message transfers started, messages relayed, message transfers aborted, messages dropped, hop count and overhead ratio all of which increased in a near linear fashion as the number of copies increased, all while seeing diminishing returns on the delivery ratio after its early peak. These metrics all coincide with a significant increase in resource usage such as bandwidth, storage and power consumption. The simulations showed that the behaviour of all maps was consistent for identical changes and therefore gave a positive indication that one set of ideal parameters could be used across the rest of the simulation. The decision was taken to use 6 copies for the rest of the simulation as well as the source version of the SaW as this provided the best combination of resource usage and performance all while creating performance patterns that could provide some insight into the map similarities. For the PROPHET the value of a 1 second for the time unit was used as the smalles values produced the best performance result and investigating the larger values of the time unit showed negligible effects on performance.

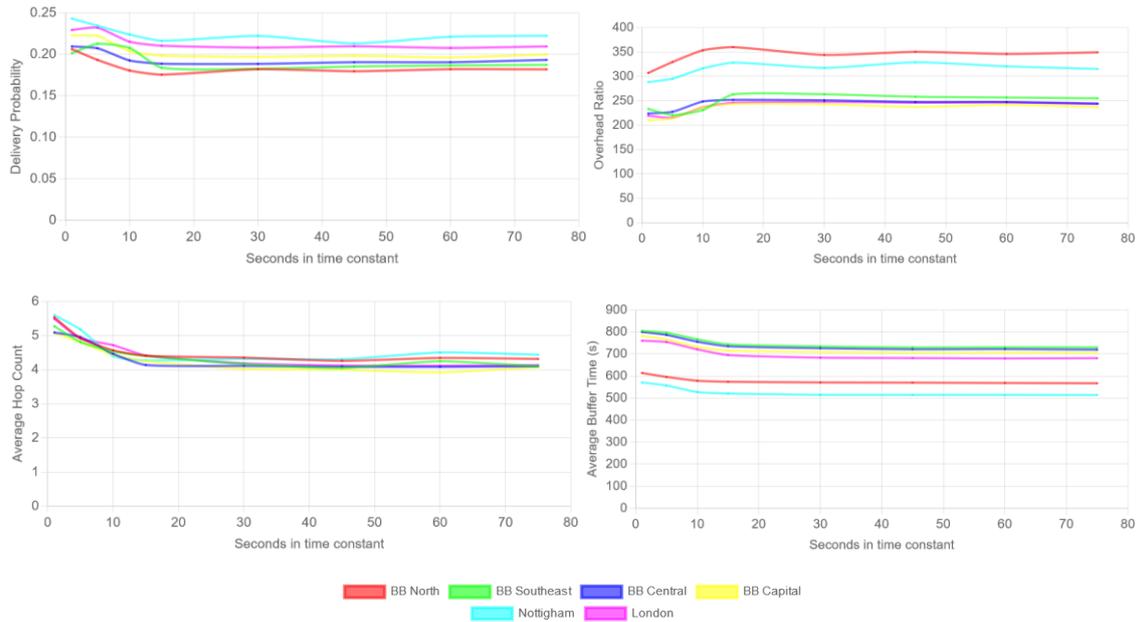

Figure 5.1: graphs showing how delivery probability overhead ratio hop count and buffer time all vary with the time unit for PROPHET

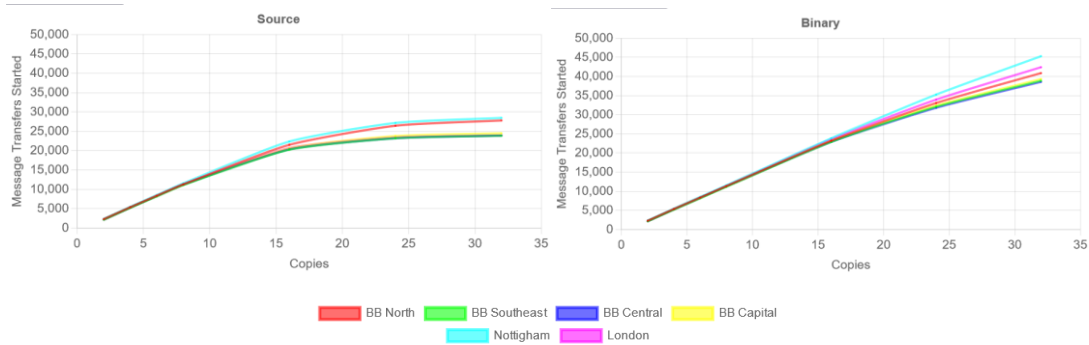

Figure 5.2: Graphs showing the amount of message transfers started for both source and binary SaW

## 5.2. Buffer Size Performance Analysis

As the buffer size increased PROPHET unsurprisingly saw increases in the number of messages started as neighbouring nodes now had more space to carry 3rd party messages to their destination. Even with this expectation, the PROPHET router, with just 1 megabyte of buffer space was attempting to send around 17,000 messages on average across all maps, before seeing a dramatic rise as buffer size increased reaching around 170,000 messages started for Barbados north and Nottingham while the other maps were hovering around 120,000, resulting in roughly 140,000 and 100,000 messages relayed respectively for the two groups for the largest buffer size of 50MB. This number is particularly extreme when considering that there were only 1,458 unique messages created per scenario meaning that on average there were more than 95 message transfers required per message for the more resource intensive maps. With over 130,000 messages dropped and an overhead ratio consistently above 100 and peaking at 350 the data clearly shows that the PROPHET router is extremely resource intensive in the scenarios provided. The delivery ratio steadily rose throughout the experiment suggesting that in spite of the plateau in relayed messages at around the 30MB buffer size mark, increases beyond this point will improve network performance. Most of the performance graphs for SaW were intuitive and the total number of relayed messages was significantly lower than PROPHET, maintaining one tight grouping in which all maps remained within 10% of the average. One interesting pattern is the spike in dropped messages between 1MB and 5MB before gradually decreasing and levelling out, this is likely to be as a result of buffers being initially so small that they aren't holding any significant number of messages to be dropped. A more in-depth look at the message buffers when recording their occupancy supports this theory as the majority of 1MB buffers are only able to hold one message as the message sizes range between 512KB and 1MB meaning that unless a node happens to receive 2 messages of size 512KB each, its buffer will be unable to store more that one message. This de-facto limit of one message per node also results in a much more unstable buffer occupancy pattern for the 1MB as shown in Figure 5.5 as the buffer occupancy for each node depends solely on the size of its one message which can range from 50-100% of the buffer size.

Unlike PROPHET, SaW had a very distinct plateau of the delivery probability after the buffer size hit about 20MB, the average latency graph produced an identical shape indicating that the increases in latency was as a result of more messages being delivered and not of a worse performing network. In general for SaW after the 20MB buffer size mark there was little to no change in performance or resource usage, while PROPHET was clearly more dependent on larger buffer sizes as performance continued to increase right up to our final increment of 50MB.

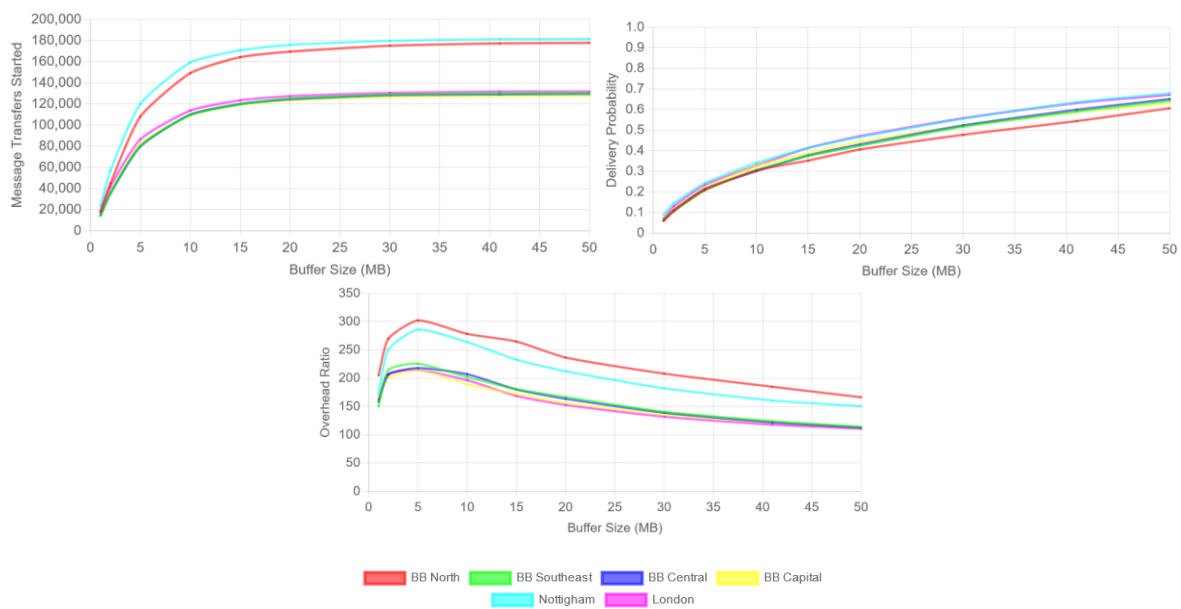

Figure 5.3: Graphs showing how buffer size influenced the delivery probability, number of message transfers started and overhead for the PROPHET router

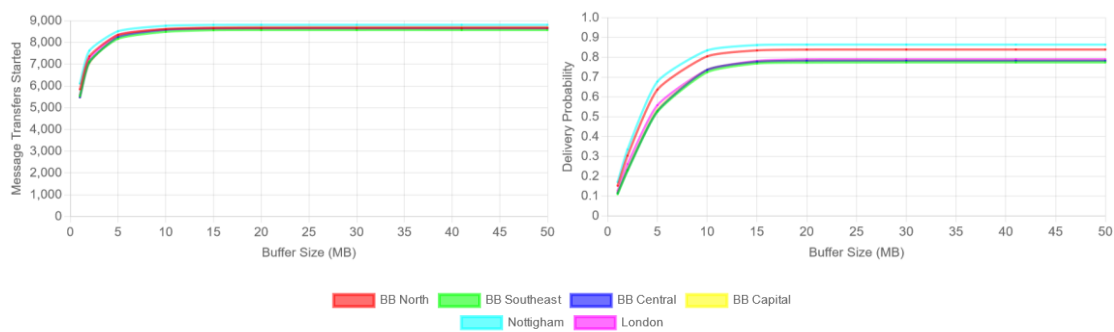

Figure 5.4: Graphs showing how buffer size influenced the delivery probability and the number of message transfers started for the PROPHET router

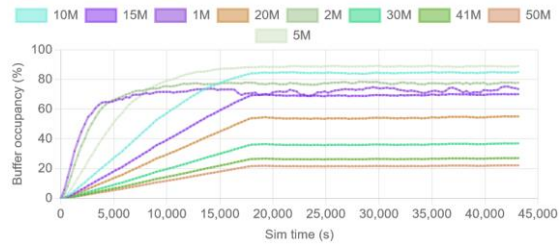

*Figure 5.5: Graphs showing how buffer size influenced the buffer occupancy for the PROPHET router*

## 5.3. Bluetooth Range Performance Analysis

As the Bluetooth range increased there was an expected increase in message transfers started as nodes would now have more opportunity to connect with other nodes farther away. For the PROPHET router the London map in particular saw a much greater rate of increase compared to all other maps likely due to the density of roads in London. Increased range also resulted in less message transfers being aborted as nodes were now able to stay within communication range for long enough periods to complete the transfer. Similar to our buffer size experiment more than 99% of transfer messages were eventually dropped. Unlike the buffer size experiment however the continuous increase in network activity did not result in similar increases in the delivery probability, and the 30m mark represented the point at which no further improvements in delivery probability occurred. Even as messages increased in this case the latency continued to drop as increased Bluetooth range simply meant that message carriers had more opportunities to come into range of either intermediary nodes or the destination node, thus moving the message along its journey more quickly. This notion is confirmed in Figure 5.6 showing a drop in average buffer time coinciding with the drop in average latency, all while the hop count gradually increased showing that increased range allowed messages to more quickly move from hop to hop on the journey to their destination. Unlike the case of PROPHET the limit of how many copies per message are created with SaW meant that increasing the range decreased the likelihood of transfers being aborted as the more successful transfers that were as a result of increased range reduced the need to attempt to send more messages and decreased the number of transfers started. This generally had a positive effect on the network as the delivery probability rose and latency decreased.

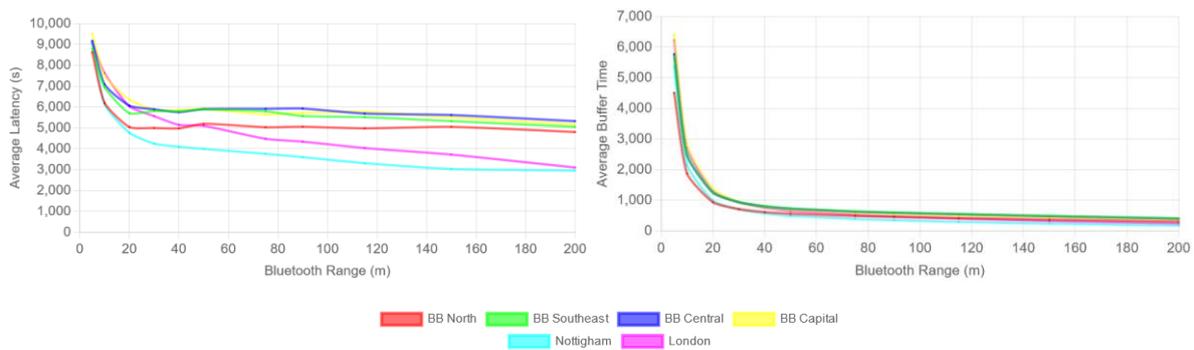

*Figure 5.6: Graphs showing the buffer time and latency simultaneously dropping as Bluetooth range increases for the PROPHET router*

## 5.4. Bluetooth Speed Performance Analysis

Similar to range, increasing Bluetooth speed was shown to have decreased the number of aborted messages transfers, in this case as a result of faster transfer speeds requiring nodes to be within contact range for a shorter period of time. In the case of SaW the percentage of messages transfers aborted for the slowest speed of 250KB/s was 5% of the total attempted, decreasing to less than 0% by the time the speed doubles to 500KB/s, reducing the total number of transfers aborted by 85% in the process. This is the only significant change that occurred as the Bluetooth speed was increased, with very little variation between maps. For the PROPHET router the increase resulted in a 90% drop in the total number of message transfers aborted between the slowest and fastest speeds. Overall the PROPHET router made use of the additional Bluetooth speed to initiate and complete more transfers which increased usage but saw no further improvements in performance.

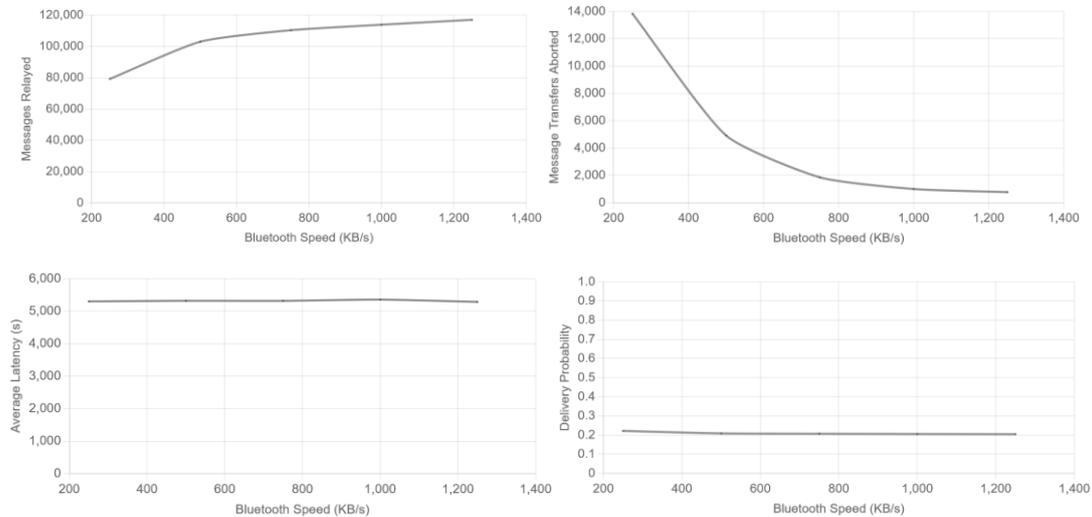

*Figure 5.7: Graph showing a cross map average of how Bluetooth speed influenced message transfers, aborted, latency and delivery probability for the PROPHET router*

## 5.5. Node Count Performance Analysis

As the number of nodes in the scenarios using PROPHET increased the number of message transfers initiated rose in an almost exponential fashion with similar trends for other resource usage metrics. This accelerated rise in resource usage was accompanied by very little significant difference in delivery probability but a similar exponential rise in dropped messages indicating that in this case having more nodes available to forward messages ultimately resulted in network overcrowding causing worse performance even though the amount of messages generated was still the same for all number of nodes. Similar to transfer speed the increase in the number of nodes present in the scenario allowed SaW to approach the mathematical limit of how many copies of a message should be forwarded, with the scenario using a limit of 6 copies and a total of 1459 messages created for the largest value of 700 nodes that theoretical limit is 8754 though it is unlikely that such a number would be reached in practice as source nodes will occasionally come across the destination before forwarding all 6 copies and additionally might become isolated or only come across nodes with full buffers preventing them from forwarding the full quota of copies. The message limit also served to reduce dropped messages as more nodes were available to forward and handle the traffic present in the network, causng the delivery probability to rise and the overhead ratio falling as nodes increased, this experiment showed one of the real strengths of the configurability of SaW.

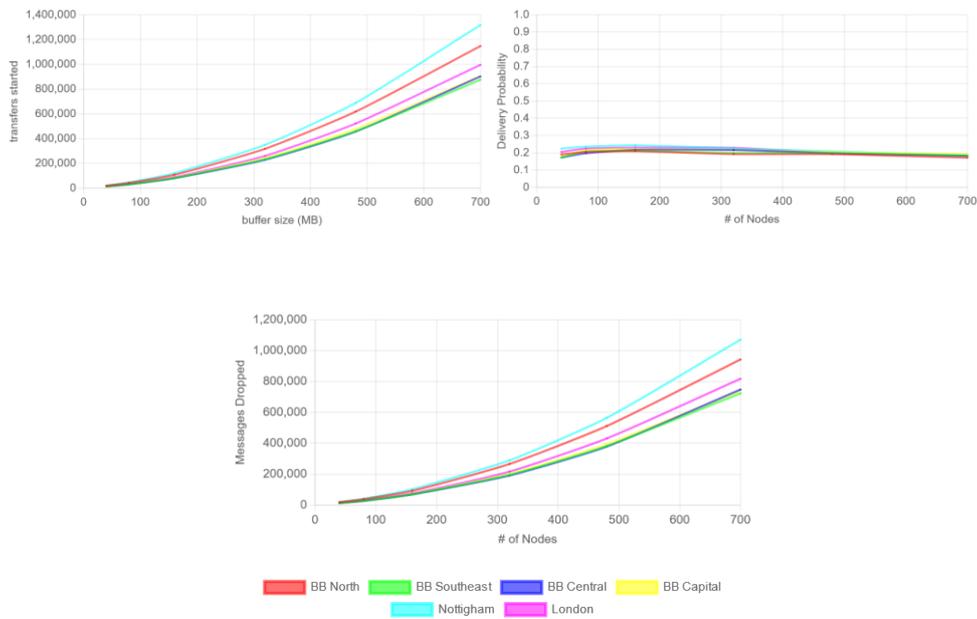

*Figure 5.8: Graphs showing how node density influenced the delivery probability, number of message transfers started and number of messages dropped for the PROPHET router*

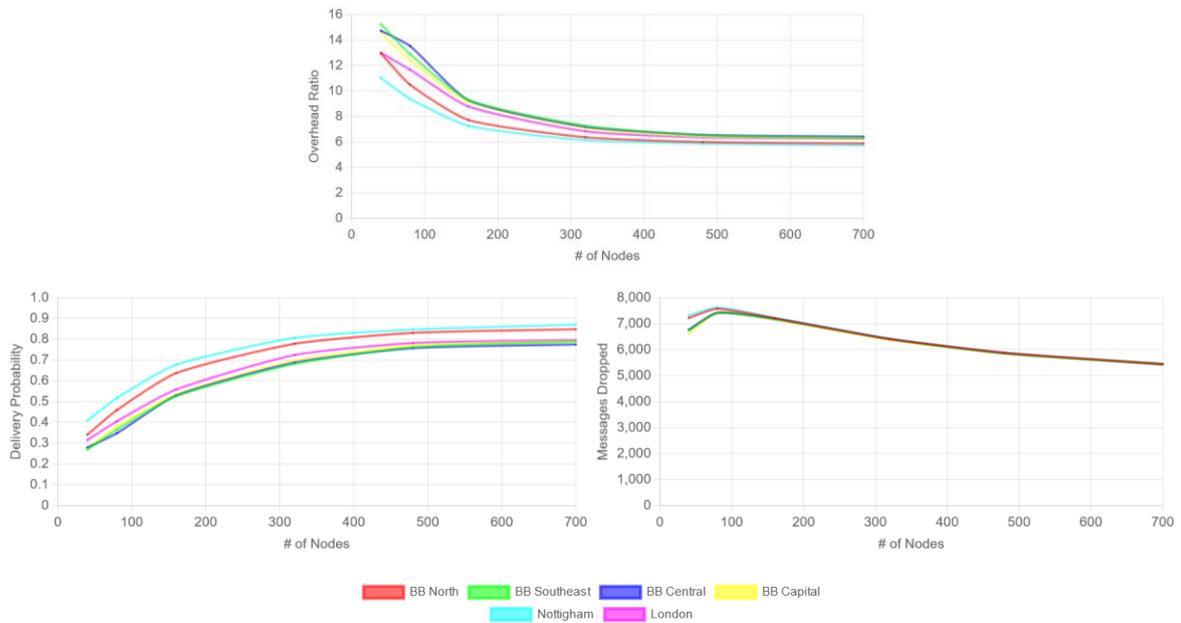

*Figure 5.9: Graphs showing how node density influenced the delivery probability, overhead and number of messages dropped for the SaW router*

The message frequency simulations were yet another indication of the PROPHET router's inefficiency and heavy resource reliance. As the number of messages created per scenario rose there was very little change in the number of message transfers started, as noted in earlier analysis PROPHET tended to use as much available resources as possible meaning there was little resource availability to initiate more transfers for these new messages and thus there would be less copies per message circulating the network. As a result Figure 5.10 shows the average delivery probability across all maps fell by over 50% whereas the overhead ratio also fell as there was a higher overall number of messages being delivered. SaW router saw message transfers started, messages relayed and messages dropped all rise in linear fashion as more messages flooded the network. The number of delivered messages however was not able to match this rise in messages and as the total continued to increase at a slower and continuously rate, the delivery probability fell significantly as a result. A constant average hop count in addition to falling average buffer times and latency indicate that the messages that were closer to their destinations were ultimately the ones being delivered before forwarding nodes had to clear buffer space.

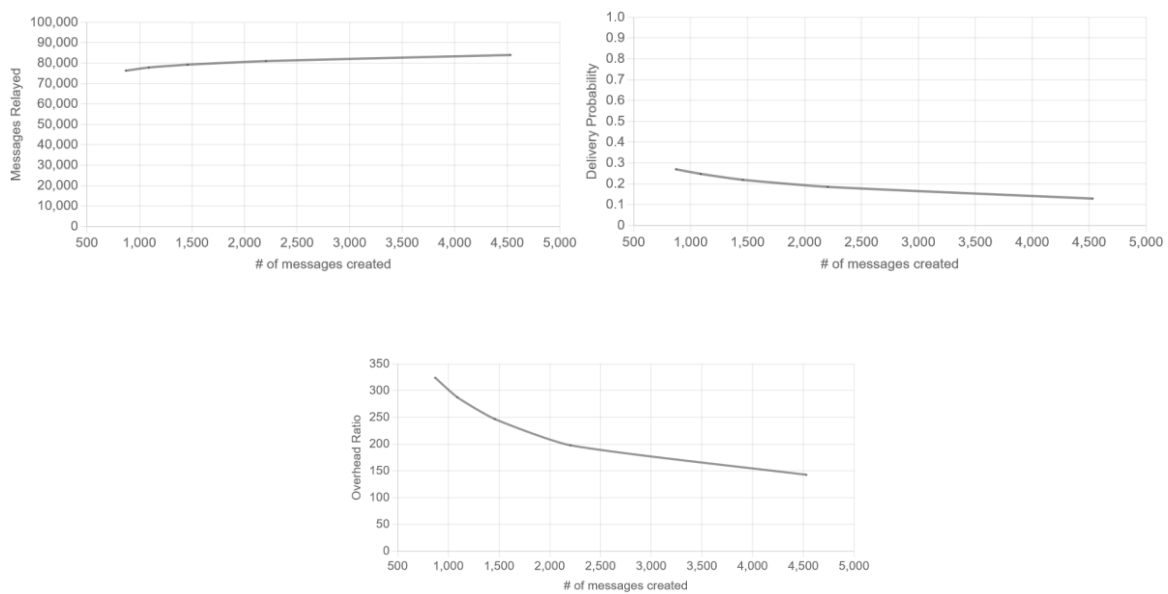

Figure 5.10: Graphs showing a cross map average of how the number of messages influenced the delivery probability, number of messages transferred and overhead for the PROPHET router

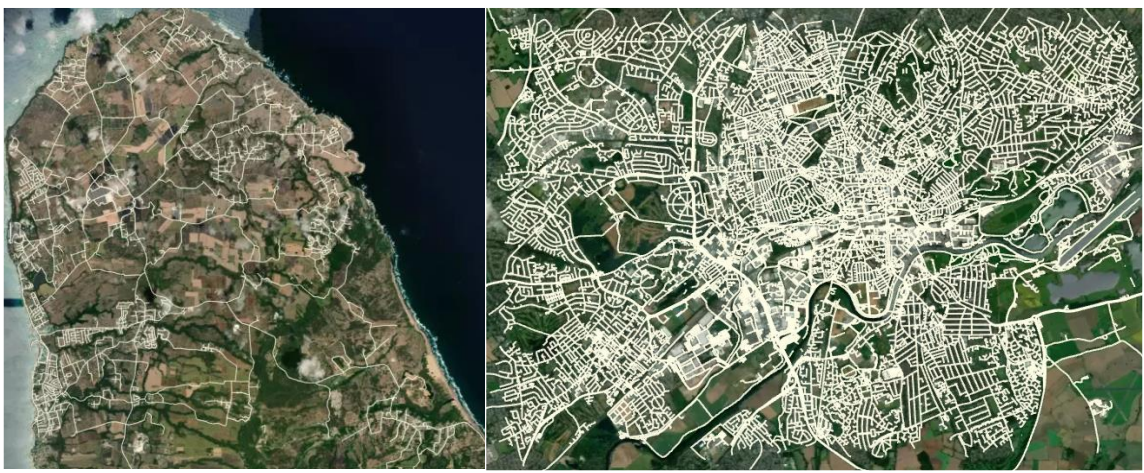

Figure 5.11: Showing the respective road densities of Barbados north and Nottingham respectively

Buffer size seemed to create a clear pattern where the Barbados north map and Nottingham were constantly very close in performance and at a noticeable distance from the other maps. This creates an interesting pairing as the northern part of the island, while home to its second largest town, has a significantly lower road density than Nottingham. This is clear in Figure 5.11 showing both road networks after tracks were removed. Given the trend of the PROPHET router being resource hungry it appears that the geography of these two maps created more opportunities for PROPHET to spread additional message copies around the network as buffer size increased. Further analysis shows that both maps have very similar contact times, close to 30% higher than the next map. Given the characteristics of the PROPHET router these higher contact times provide more opportunity for nodes to exchange more messages driving that increase seen in resource usage. A closer look at the Bluetooth range contact times in Figure 5.14 shows that as the range increases the London map has consistently higher contact time instances than Barbados north, while Nottingham continued to rise producing a more intuitive end result of both Nottingham and London having higher contact times though Nottingham remained more than 30% higher than London at it's peak contact time length. The results of this shift in contact time durations means that at lower contact times the Barbados north map produces a network performing more closely associated with high density maps, while the London map's performance was closer to the less densely populated maps, however as the range increased the performance of these two gradually went in opposite directions and they essentially swapped places. We can see these effects taking place in Figure 5.12 where the graphs for both latency and message transfers started both saw an initial phase where the Barbados north and Nottingham plots began drifting away from the remaining plots before Barbados north drifted back towards the core group and London drifted away and closer to Nottingham. As was the case in our general analysis the increased activity seen by the PROPHET router on the Barbados north and Nottingham maps resulted in no significant improvements of their delivery probability over other maps. The map characteristics had a reverse effect on SaW, with its resource usage being well controlled and staying almost identical across maps as the range increased, the delivery probability though once again revealed a similar pattern of behavior for Barbados north and Nottingham. For the remaining 3 experiments there was a trend consistent with our analysis as the graph plots for PROPHET resource usage tended to be split into 2 groups one with Barbados north and Nottingham and the other maps in the second group. On occasions there would be relatively significant space between the two breakaway maps but with both clearly being separated from the second group. On the other hand SaW continued to be diligent in it's low resource usage for the better connected maps but instead seeing better delivery probabilities. One reason for the behavior of these maps being so different from the rest, especially in the case of Barbados north which has a significantly more sparse road network, is that the sparse nature accompanied by smaller areas of more dense road networks create a network of DTN "islands" that stay well connected with each other for a period of time before travelling off to the next island to deliver any messages destined for that area. This is also supported by the performance of SaW vs PROPHET where SaW was able to make much better use of increased node density and buffer space because they both increased the total buffer availability in the network meaning that nodes can now travel from island to island receiving copies of messages yet to reach their copy limit and move onward to potentially find the recipient. In the case of PROPHET however these nodes will be quickly overwhelmed moving from island to island and constantly receiving more messages, prompting them to drop messages just as frequently as they are received therefore having no impact on the delivery probability. While the map of Nottingham had a significantly more dense network of roads there were several large pockets of space that could serve to create a similar island like network, but additionally as the Bluetooth range increases the more dense road network allows for each incremental increase to discover fresh contacts, whereas the nodes in the smaller "islands" within Barbados north will find that after the range expands to cover the area of that "island" there will be very little neighbouring nodes left to discover, hence the fall-off in contact time instances compared to the rest of the maps.

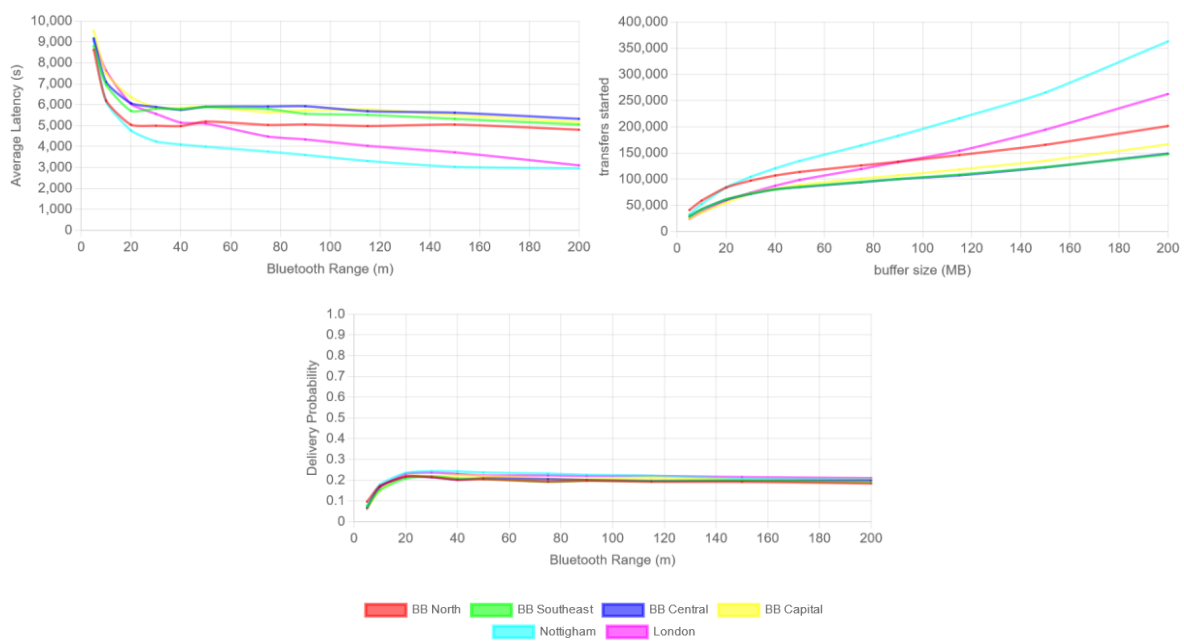

Figure 5.12: Graph showing latency and message transfers started for the PROPHET router as Bluetooth range increases

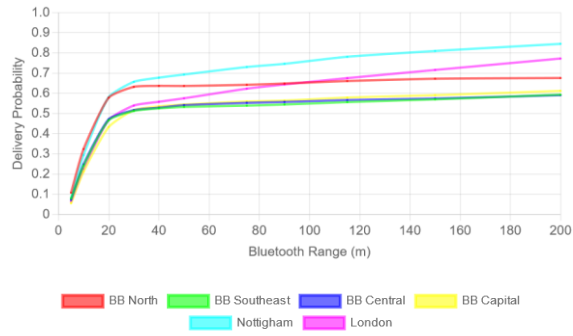

*Figure 5.13: Delivery probability for SaW as Bluetooth range increases*

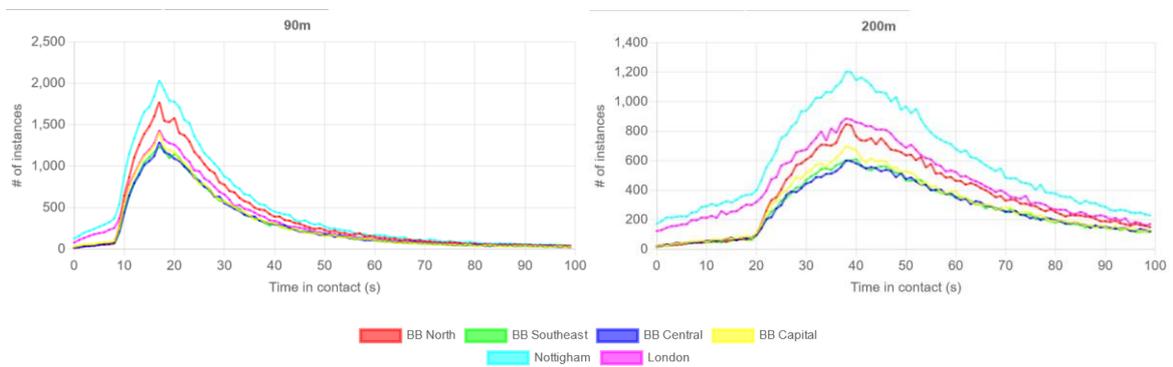

*Figure 5.14: The contact time for all maps at the Bluetooth ranges of 90m and 200m*

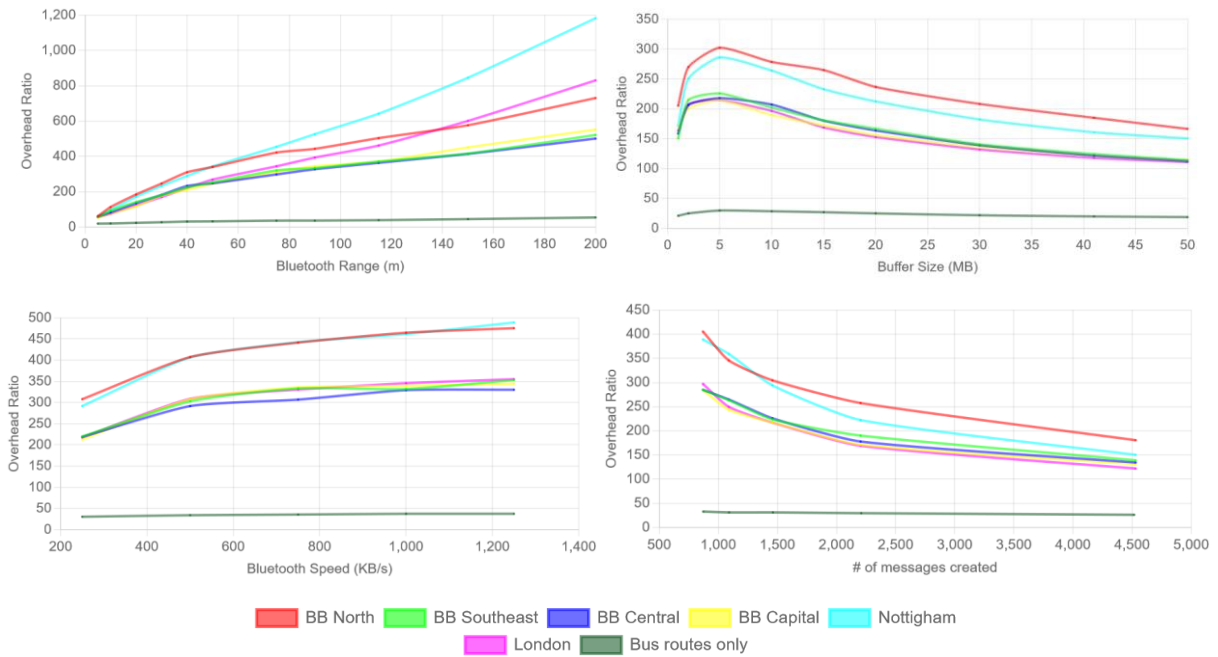

*Figure 5.15: Graphs showing the overhead for the bus routes compared to all other maps for various experiments*

When running the simulations on the map consisting of the 8 bus routes there were mixed results for both routers, the predictability introduced by having bus nodes travel back and forth on their respective routes seemed to do little to assist the prediction based algorithm of PROPHET. Compared to the main maps the delivery probability was either on par with, or below for PROPHET though one clear upside is that the limited amount of nodes available meant that the total amount of message transfers was also reduced significantly resulting in a significant decrease in the overhead ratio as shown in 5.15.

For SaW the delivery probability consistently lagged behind the main maps, while having a comparable number of total transfers meaning that the overhead ratio was much higher than the main maps. For both routers the increases in buffer size created the biggest boost in delivery probability, as the limited number of nodes in the network were now able to receive more message copies. The connectivity of the network was unique when compared to the main maps, beginning with the contact times which produce a similarly shaped graph to the other maps but was a lot less smooth, as Figure 5.16 shows the bus routes exhibit numerous spikes as the graph for Barbados north saw a rapid, smooth decline in how many instances there are of the longer contact times after an initial sharp drop, the graph makes a much more prolonged decent that was consistent in shape with the others. One thing to note as it related to the contact times is that the nodes on a bus route will often be travelling on the same path as each other in the same direction and therefore often times manage to stay within connection range for longer than a typical pair of nodes who would be on different paths. When also considering the overlap occurring between routes this creates a temporary merging point between multiple routes that can often result in increased node traffic in both directions and thus increase the likelihood that two nodes might form a connection for the full distance of this overlap. This overlap could potentially explain why even after taking the average across 3 different runs there were very distinct spikes for some lengths of contact time. The experiment that gave the most insight into the performance deficit when analysing the bus routes was the buffer size experiment which showed that for both routers increasing the buffer size significantly boosted delivery probability as shown in Figure 5.17. As we discussed previously in Section 5.2.3, the general pattern of the delivery probability for PROPHET was to continuously rise as the buffer size increased making the result for the bus route experiment in line with more general expectations. However in the case of SaW it was clear that the increases in delivery probability were more specific to the map layout as the rate of increase was more gradual compared to the main maps and continued to rise beyond the point at which the other maps had their plateau.

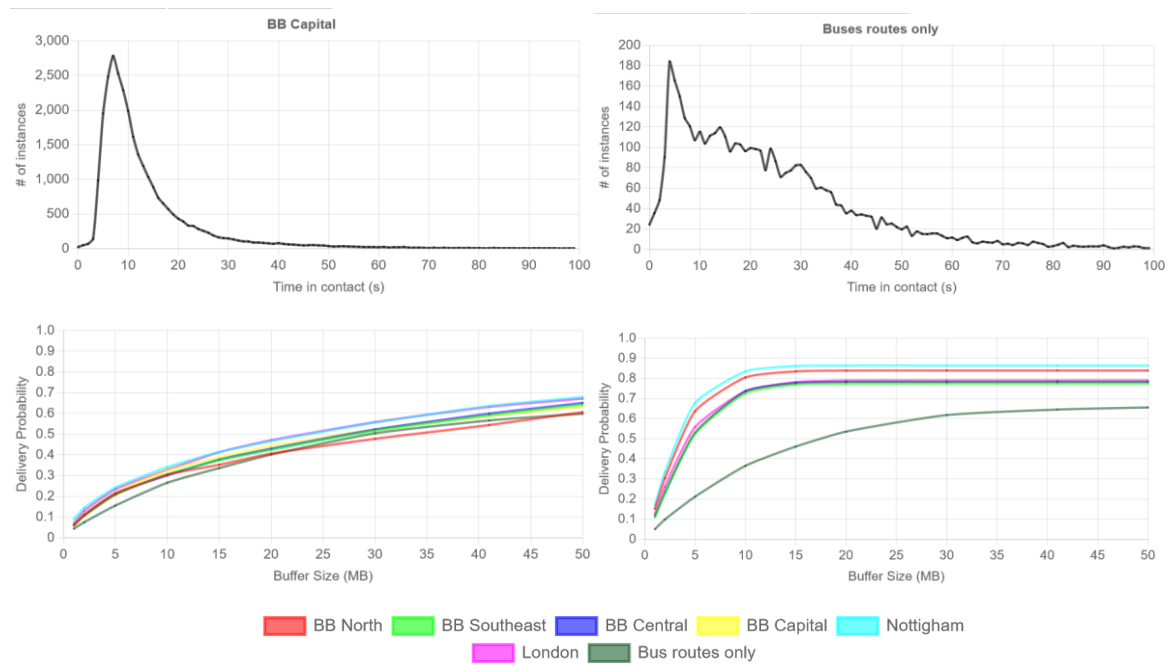

*Figure 5.17: How the delivery probabilities for the route only map compares to the other maps for both PROPHET and SaW as buffer size increases*

# 6. Summary and future work

The data gathered in our experiment gave strong suggestions that our initial goal of finding similarities between developing countries, in this case Barbados and their more developed counterparts was within sight. There were the many graphs that showed there was a strong link between the performance of the Barbados north and

Nottingham maps and how they responded to changes in environment. Another potentially more subtle takeaway is that with the larger group containing London and the other three maps from Barbados there was also a consistent trend of these maps performing very similar, meaning that there can also be a strong case made for their similarities as it relates to the type of networks they create.

We aim to carry out a lot of more work in order to draw more concrete conclusions such as being able to accurately emulate realistic node movement and density has the potential to have a much more profound effect on how the networks perform. By only taking into account the road lines, simulations miss out on opportunities to emulate things like one way streets, speed limits and also how vehicular traffic can affect the speed of travel. Therefore, one clear path forward to iterate on this research is to collect and apply real world data gathered from the field in order to use the working day model that provides the ability to define work places, and other points of interest, and has the notion of day/night cycles along with many other parameters that map to aspects of day to day life that would have a significant baring on how nodes behave. Similarly, in order to provide a more accurate description for how buses might move, future work could include building on top of the already available bus control system to keep track of a fleet of finite buses, control when buses leave the various terminals and managing how to respond to delays that might be caused by the busier periods created by the *workingday* model. Overall the project can be considered a success taking all limitations into account as it has done enough to serve as a platform for further research into the topic investigating theories proposed and building on the data collected. We envisage that we will investigate the deployment of the state of the art AI based geo temporal spatial DTN approaches such as CognitiveCache [29] and SmartCharge [30] , incorporate them in the MODiToNeS platform [26] , and deploy them in the next generation Barbados ad-hoc network services inspired by our deployments in highly challenging environments [27] [28].